\documentclass[twocolumn,letterpaper,amsmath,amssymb,floatfix,aps,superscriptaddress]{revtex4}
\usepackage{blindtext}
\usepackage{graphicx, xcolor}
\usepackage{rsfso}
\usepackage{mathrsfs}
\usepackage{amsmath, dcolumn}
\usepackage{amsmath}
\usepackage{mathtools}
\usepackage{bm}
\usepackage[breaklinks=true,colorlinks=true,linkcolor=blue,urlcolor=blue,citecolor=blue]{hyperref}
\usepackage{hyperref}
\usepackage{ulem}

\begin{document}
\title{Driven Polymer Translocation through a Nanopore from a Confining Channel}

\author{Soheila Emamyari}
\affiliation{School of Quantum Physics and Matter, Institute for Research in Fundamental Sciences (IPM), Tehran 19538-33511, Iran}

\author{Jalal Sarabadani}
\email{jalal@ipm.ir}
\affiliation{School of Quantum Physics and Matter, Institute for Research in Fundamental Sciences (IPM), Tehran 19538-33511, Iran}

\author{Ralf Metzler}
\affiliation{Institute of Physics and Astronomy, University of Potsdam, 14476 Potsdam, Germany}
\affiliation{Asia Pacific Centre for Theoretical Physics, Pohang 37673, Republic of Korea}

\author{Tapio Ala-Nissila}
\affiliation{Department of Applied Physics and QTF Center of Excellence, Aalto University, P.O. Box 15600, FIN-00076 Aalto, Espoo, Finland}
\affiliation{Interdisciplinary Centre for Mathematical Modelling and Department of Mathematical Sciences, Loughborough University, Loughborough, Leicestershire LE11 3TU, United Kingdom}

\begin{abstract}
We consider the dynamics of pore-driven polymer translocation through a nanopore to semi-infinite space when the chain is initially confined and equilibrated in a narrow channel. To this end, we use Langevin dynamics (LD) simulations and iso-flux tension propagation (IFTP) theory to characterize local and global dynamics of the translocating chain. The dynamics of the process can be described by the IFTP theory in very good agreement with the LD simulations for all values of confinement in the channel. The theory reveals that for channels with size comparable to or less than the end-to-end distance of the unconfined chain, in which the blob theory works, the scaling form of the translocation time depends on both the chain contour length as well as the channel width. 
Conversely, for a very narrow channel the translocation time only depends on the chain contour length and is similar to that of a rod due to the absence of spatial chain fluctuations.
\end{abstract}


\maketitle

\section{Introduction}\label{intro}

Since the seminal experimental studies by Bezrukov {\it et al.}~\!\cite{Bezrukov} and Kasianowicz {\it et al.}~\!\cite{KasiPNAS1996}, and the theoretical work by Sung and Park~\!\cite{SungPRL1996}, a huge body of theoretical \cite{Muthukumar_book,Tapio_review,jalalJPCM2018,MilchevJPCM2011,SungPRL1996,MuthukumarJCP1999%
,Chuang2001,SlaterRatchet,Kantor_PRE2004,tobias,roya,GrosbergPRL2006,Luo2006,SakauePRE2007%
,Luo2007,aksimentievNanolett2008,slaterPRE2009,kaifu,kaifu1,kaifu2,kaifu3,SakauePRE2010%
,slaterPRE2010,rowghanian2011,golestanianPRL2011,RalfJCP2011,SakauePRE2012%
,ikonen2012a,ikonen2012b,slaterJCP2012,Luo_SoftMatter_2012,hamidJCP2013,jalalJCP2014%
,MichelettiMacLett2015,Aniket_2015,jalalJCP2015,FaloPRE2015,menais_2017,JalalEPL2017%
,MichelettiPNAS2017,jalalSciRep2017,menais_2018,jalalPolymers2018,jalalPolymers2019%
,Aniket_2020,jalalJPCM2020_1,jalalJPCM2020_2,jalalPRR2021,JalalPRR2022,jalalJPCM2023,JalalPRR2023} and experimental \cite{mellerPRL2001,BrantonPRL2003,Storm2003,MellerBiophysJ2003,Keyser_2006,RadenovicNanoLett2007,
Keyser_2009,RadenovicNanoscale2012,LeeChemComm2012,RadenovicNutreNanoTech2013,Bulushev_2015} works have appeared to explain the underlying physical mechanisms during the process of polymer translocation (PT) through a nanopore. 
In addition to being of great theoretical interest as a dynamical nonequilibrium process, PT has a wide range of applications such as RNA transport through nuclear membrane pores~\!\cite{nuclear}, transfer of genes between bacteria~\!\cite{bact}, using ratchets for filtration of polyelectrolytes~\!\cite{SlaterRatchet}, and drug delivery and DNA sequencing~\!\cite{drug_DNA,Branton_Naturebio,Deamer_Naturebio}.

There are many variations of the PT process that have been studied to date. The translocation process can be either unbiased~\!\cite{Chuang2001,Luo2006,slaterPRE2009,slaterPRE2010} or driven~\!\cite{SakauePRE2007,rowghanian2011,jalalJPCM2018}. For the driven case, several different scenarios exist, for example, in the end-pulled case an external driving force can be applied to the polymer head monomer~\!\cite{JalalEPL2017} by an atomic force microscope (AFM) \cite{RitortJPCMatt2006}, or by magnetic or optical tweezers \cite{Keyser_2006,Keyser_2009,Bulushev_2014,Bulushev_2015,Bulushev_2016}. The driving force can be localized inside the pore (pore-driven PT)~\!\cite{SakauePRE2007,rowghanian2011,jalalJCP2014}, or it can be an effective force due to interaction of the {\it trans}-side subchain with ambient active rods \cite{jalalPRR2021} or chaperones \cite{tobias,roya}.  In the pore-driven case where the driving force acts on the monomers inside the nanopore, the driving force can be alternating and the nanopore can be flickering~\!\cite{jalalJCP2015,golestanianPRL2011}. Both the pore-driven and end-pulled cases can be described by the iso-flux tension propagation (IFTP) theory~\!\cite{SakauePRE2007,rowghanian2011,jalalJCP2014,jalalJPCM2018}.
In the unbiased or weakly driven PT, the entropic force resulting from fluctuations in the spatial configurations of the chain (entropy) plays an important role in the dynamics of the process and a full theoretical understanding of PT in this limit is still missing~\!\cite{SungPRL1996}. 

An interesting special case of PT is when there is a significant entropic force due to geometric confinement either on the {\it cis} or the {\it trans} side. In the latter case of polymer trapping, an external driving force is needed to overcome entropic repulsion due to confinement.
In Refs.~\!\onlinecite{RalfJCP2011} and \onlinecite{JalalPRR2022} driven PT of a polymer from a semi-infinite {\it cis} side into a 
{\it trans} side channel was investigated. In the case where the polymer is initially trapped, it can spontaneously escape the confinement or a confining potential well {~\!\cite{Muthukumar_PRL_2001,Luijten_Nano_Letters,Luo_PRE_2009}.} This process can be further facilitated by external driving. In Ref. 
\onlinecite{Slater_2015} a pore-driven PT process was considered, where the polymer was confined in a channel on the {\it cis} side and then it translocated to a semi-infinite compartment through a pore in the middle of the channel. Using Langevin Dynamics (LD) simulations and tension-propagation arguments, the authors were able to characterize the dependence of PT on the channel diameter and obtain an analytic scaling form for the average translocation time $\tau$.

\begin{figure*}
\includegraphics[width=0.99 \textwidth]{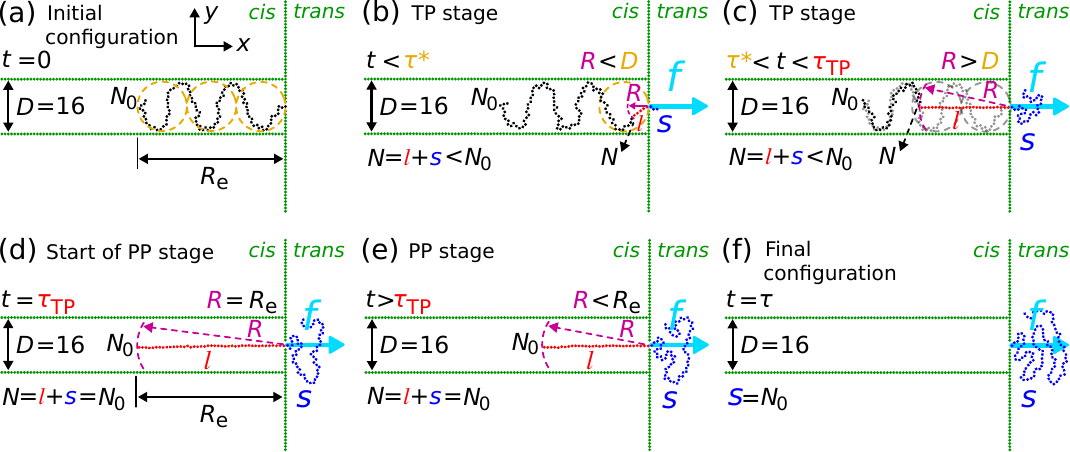}
\caption[]{(a) Polymer configuration inside the channel on the {\it cis} side just after equilibration at time $t=0$. The contour length of the polymer is $N_0=100$ and $D=16$ is the channel width. $R_{\textrm{e}}$ is the end-to-end distance of the polymer at $t=0$ which is a function of $D$ and $N_0$. The orange dashed circles show the blobs whose diameters are equal to the channel width here. The channel walls and the membrane are parallel to the $x$ and $y$ directions, respectively. The center of the pore is located in the middle of the channel at the origin $(0,0)$. 
(b) Configuration of the polymer during the tension propagation (TP) stage when $s$ monomers have already translocated to the {\it trans} side. 
$f$ is the external driving force (denoted by the horizontal cyan vector) acting on the monomer(s) inside the pore in the positive $x$ direction towards the {\it trans} side. 
$R$ (denoted by the violet vector) is the location of the tension front at time $t$, and $l$ is the number of red monomers under the tension on the {\it cis} side. Here, $R$ is smaller than $D$ for $t<\tau^*$, where $\tau^*$ is the time at which $R = D$. $N$ is the number of monomers that have been influenced by the tension so far, i.e., $N= l + s < N_0$.
(c) Polymer configuration during the TP stage when the tension front has not yet reached the last monomer of the polymer while $R>D$, i.e., for $\tau^* < t<\tau_\text{TP}$. Here, the chain in gray color shows the polymer configuration at $t=0$ just to compare with the chain configuration at time $t$.
Panel (d) is the same as (c) but the tension front has just reached the last monomer of the polymer at $t=\tau_\text{TP}$ when $R=R_{\textrm{e}}$, and as all monomers are under tension, $N= l + s = N_0$.
(e) The post-propagation stage of the translocation process at $t>\tau_\text{TP}$, where all monomers of the {\it cis} side are moving towards the pore.
(f) The final configuration of the polymer just at the end of the translocation process at $t=\tau$ in which $s=N_0$.
}
\label{fig-schematic}
\end{figure*}

In this paper we expand upon the work in Ref. \onlinecite{Slater_2015} by performing a detailed analysis of the pore-driven PT process. In analogy to their system, we drive a flexible polymer through a nanopore where the polymer starts in a long confining channel on the {\it cis} side. It is forced through a nanopore by an external force that affects the monomers inside the pore (cf. Fig.~\!\ref{fig-schematic}). We characterize the dynamics of this process by using the full IFTP theory and scaling analysis for driven PT, and derive analytic expressions for $\tau$ as a function of the chain length, driving force, confinement dimension, and effective channel friction generalizing the results of Ref. \onlinecite{Slater_2015}. Numerical data from the LD simulations are then compared with theory for the relevant physical quantities. The main interest here is the influence of channel width $D$ and driving force $f$ on PT dynamics.

In the high-force limit in which the entropic force can be ignored, regardless of the width of the channel, the IFTP theory predicts that  $\tau$ scales with the force as $\tau \propto f^{\beta}$ with $\beta = -1$. However, the results of the LD simulations for a fixed large $f$ show that $\vert\beta\vert < 1$ and it depends on $D$. It increases with increasing channel width due to the growth of fluctuations from $-0.925 \pm 0.008$ to $-0.900 \pm 0.003 $. For $D > R_{\textrm{e}}$, where $R_{\textrm{e}}$ is the end-to-end distance of the unconfined chain, the force exponent settles to $-0.900 \pm 0.003$ as the polymer is very weakly confined and fluctuations in the spatial polymer configurations do not increase.
There are experimental results that have visualized the spatial conformations of a DNA inside a confining micro or nanochannel~\!\cite{Tegenfeldt_ABC2004,Tegenfeldt_PNAS2004,%
Tegenfeldt_PNAS2005,Tegenfeldt_book,Tegenfeldt_CSR2010}, so we  expect that comparing our results with the experimental observations leads to a better understanding of this translocation process.

The structure of the paper is as follows. In Sec.~\!\ref{model} the simulation model is explained in detail. Then, the IFTP theory is introduced in Sec.~\!\ref{theory}. Results that include waiting time, translocation time, mean squared displacement and monomer number density are presented in Sec.~\!\ref{results}. Section~\!\ref{conclusion} contains our summary and conclusions.

\section{Simulation Model}\label{model}

For the numerical LD simulations we have used the LAMMPS software package \cite{lammps}. The system under consideration is two-dimensional composed of a linear self-avoiding, fully flexible polymer initially on the {\it cis} side channel of width $D$ as depicted in Fig.~\!\ref{fig-schematic}(a). At the end of the channel, there is a membrane with a pore in the middle of the channel through which driven PT takes place.
The membrane and channel walls are made up of fixed particles each of size $\sigma$ that are located next to each other at a distance of $\sigma$ (green beads in Fig.~\!\ref{fig-schematic}). 
The pore diameter is $2\sigma$ and the channel length is much larger than the polymer contour length. The simulation box dimensions in the $x$ and $y$ directions are $L_x=L_y=210 \sigma$.
The polymer is composed of $N_0 $ monomers (black beads in Fig.~\!\ref{fig-schematic}(a)) and is modeled by a bead-spring chain~\!\cite{grest}.
The consecutive bonded monomers interact with each other via a sum of Weeks-Chandler-Anderson (WCA)~\!\cite{wca} and finitely extensible nonlinear elastic (FENE) potentials. The repulsive WCA potential is
\begin{equation}
U_\text{WCA}(r) = \left\{
\begin{array}{ll}
U_\text{LJ}(r)-U_\text{LJ}(r_\text{c}), & \text{~~if $~r\leq r_\text{c}$};\\
0, &\text{~~if $~r> r_\text{c}$},
\end{array}\right.
\end{equation}
where $r$ is the monomer-monomer distance, $r_{\text c} = 2^{1/6}\sigma$ is the cutoff radius, and $U_\text{LJ}(r)$ is the Lennard-Jones (LJ) potential
\begin{equation}
U_\text{LJ}(r)=4\varepsilon\left[\left(\frac{\sigma}{r}\right)^{12}-\left(\frac{\sigma}{r}\right)^{6}\right],
\end{equation}
with $\varepsilon$ and $\sigma$ as the depth of the potential and the monomer size,
respectively. The FENE potential that connects the consecutive monomers is given by 
\begin{equation}
U_\text{FENE}(r)=-\frac{1}{2}kR_0^2 \ln \bigg[1-\left(\frac{r}{R_0}\right)^2\bigg],
\end{equation}
where $k$ and $R_0$ are the effective spring constant and the maximum extent of the bond between the neighboring monomers, respectively.
All interactions between the non-bonded monomers of the polymer, monomers and the membrane wall particles, and between monomers and the channel particles, are governed by the repulsive WCA potential.

Using the LD simulation method, the equation of motion for the position vector of the $i$th monomer, $\vec{r}_i$, is given by 
\begin{equation}
M\ddot{\vec{r}}_i=-\eta \dot{\vec{r}}_i-\vec{\nabla}U_{i}+\vec{\xi}_i(t),
\end{equation}
where $M$, $\eta$ and $U_i$ are the monomer mass, the friction coefficient, and the sum of all interaction potentials experienced by the $i$th monomer, respectively. $\vec{\xi}_i(t)$ is thermal white noise vector at time $t$ with zero mean $\langle \vec{\xi}_i(t)\rangle=\vec{0}$ and correlation $\langle\vec{\xi}_i(t)\cdot\vec{\xi}_j(t')\rangle=4\eta k_\text{B}T\delta_{ij}\delta(t-t')$. $k_\text{B}$ and $T$ are the Boltzmann constant and the temperature, and $\delta_{ij}$ and $\delta(t-t')$ are Kronecker and Dirac delta functions, respectively.

$M$, $\sigma$ and $\varepsilon$ are chosen to be the units of mass, length and energy, respectively.
$\sigma=1$ is the diameter and $M=1$ is the mass of each particle (monomers of the polymer or particles of the membrane and the channel walls). 
The values of the energy unit, solvent friction coefficient and temperature are $\varepsilon=1$, $\eta=0.7$ and $k_\text{B} T=1.2$, respectively.
The simulation time step is $dt=0.001\tau_0$ in which $\tau_0=\sqrt{M\sigma^2/\varepsilon}$ is the simulation time unit. In addition, the maximum allowed distance between each two connected monomers is $R_0=1.5$, and $k=100$ (unless otherwise mentioned) is the spring constant in the FENE potential. The contour length of the polymer in LJ units is $N_0=100$.

The size of each bead in our model corresponds to the Kuhn length of a single-stranded DNA and is approximately $1.5$~nm which is about the size of three nucleotides. As the mass of a nucleotide is about $312$~amu, then the mass of each bead in our simulations is about $936$~amu with the interaction strength of $3.39\times 10^{-21}$~J at room temperature ($T=295$~K). 
Consequently, the time and the force units are obtained as $32.1$~ps and $2.3$~pN, respectively. 
Given the pore thickness $1\sigma=1.5$~nm (see Fig.~\!\ref{fig-schematic}), and the fact that the electric charge of each bead is equivalent to that of three single nucleotides (each with an effective charge of $0.094${\it e} \cite{BrantonPRL2003,unitCharge2}), an external driving force of $f=4$ corresponds to a voltage of $306$~mV across the pore \cite{voltage}.

Before the PT process starts,  one head bead of the polymer is fixed in the pore while the rest of the chain is carefully equilibrated within the channel. 
After equilibration, the fixed head monomer is released and  the pore driving force $f$ is switched on and the translocation process starts. It ends when the last tail bead has moved to the {\it trans} side, and the time taken defines the PT time $\tau_i$ for each attempt $i$. 
To obtain sufficient statistics, data are averaged over 1000 independent PT events.
         

\section{Theory}\label{theory}

In Fig.~\!\ref{fig-schematic}(a) we show a typical equilibrium state of the system just before the translocation process ($t=0$) for a channel width $D=16$. The end-to-end distance of the polymer at time $t=0$ is denoted by $R_{\textrm{e}}$ and it is a function of $D$ and $N_0$. 
More details about the scaling form of $R_{\textrm{e}}$ at $t=0$ can be found in Appendix~\!\ref{Poly_initial_conf}. Moreover, the orange dashed circles represent blobs~\!\cite{deGennes_Book} whose diameters are equal to the channel width here.

As panel (b) in Fig.~\!\ref{fig-schematic} shows, the external driving force $f$ acts on the monomer(s) inside the pore. 
Here, $s$ is the number of beads (in blue color) that have translocated to the {\it trans} side. The subchain on the {\it cis} side consists of two parts, a part with $l$ monomers under tension (in red color) and an equilibrium part with the monomers that have not yet been affected by tension and have zero average velocity (in black color). The boundary between the two parts, denoted by $N$, defines the tension front position in the chain at a distance of $R$ (violet arrow) from the nanopore. 
Figure~\!\ref{fig-schematic}(b) presents the system configuration for $R<D$ at $t<\tau^*$, where $t=\tau^*$ is the time at which $R=D$.
As time passes with the propagation of tension along the backbone of the subchain on the {\it cis} side, $R$ grows such that $R>D$ for $t>\tau^*$ (see Fig.~\!\ref{fig-schematic}(c)). 
The stage where tension still propagates along the backbone of the polymer and has not reached the chain end on the {\it cis} side, is called the tension propagation (TP) stage. Panels (b) and (c) in Fig.~\!\ref{fig-schematic} present the TP stage of the translocation process. 
The TP time $\tau_\text{TP}$ is the time at which the tension reaches the end of the polymer chain in the {\it cis} side. 
As shown in panel (d) of Fig.~\!\ref{fig-schematic}, tension has just reached the last monomer of the subchain on the {\it cis} side at time $t=\tau_\text{TP}$, and $R = R_{\textrm{e}}$ applies.
After that, in the post-propagation (PP) stage for $t>\tau_\text{TP}$, all monomers of the subchain on the {\it cis} side move towards the pore as shown in Fig.~\!\ref{fig-schematic}(e). 
Finally, panel (f) in Fig.~\!\ref{fig-schematic} shows the final configuration of the system at the end of the translocation process at $t=\tau$, where $\tau$ is the average translocation time and $s=N_0$.

To model the driven PT process explained above, we use the well-established IFTP theory~\!\cite{SakauePRE2007,rowghanian2011,jalalJCP2014,jalalJPCM2018}. We express all quantities in dimensionless units as identified by a tilde. 
They are denoted as $\tilde{Z} \equiv Z / Z_{\textrm{u}}$, where the units of length, time, velocity, monomer flux, friction and force are indicated by the denominator, as $s_{\rm u} \equiv \sigma$, $t_{\rm u} \equiv \eta \sigma^2 / (k_{\rm B} T) $, $v_{\rm u} \equiv \sigma / t_{\rm u}$, $\phi_{\rm u} \equiv k_{\rm B} T / (\eta \sigma^2 ) $, $\Gamma_{\rm u} \equiv \eta$ and $f_{\rm u} \equiv k_{\rm B} T / \sigma$, respectively. All parameters in LJ units do not have a tilde.

From the IFTP theory the translocation process can be described by solving an equation of motion for the translocation coordinate $\tilde{s}$. To this end, one needs to know the tension force $\tilde{\mathbb{F}} ( \tilde{x} , \tilde{t} ) $ at distance $\tilde{x}$ from the pore on the {\it cis} side. 
The force-balance equation for a differential element of the mobile part of the polymer chain on the {\it cis} side is $ d \tilde{\mathbb{F}} (\tilde{x}' , \tilde{t}) =  - \tilde{\phi} (\tilde{t}) d \tilde{x}' $. Here, $\tilde{\phi} (\tilde{t}) = {d \tilde{s}}/{d \tilde{t}}$ is the monomer flux and iso-flux means that the value of the monomer flux in the mobile domain is constant. Integrating the above force-balance equation from the pore located at $\tilde{x}' = 0$ to the distance $\tilde{x}' = \tilde{x}$, the tension force at the distance $\tilde{x}$ can be written as $ \tilde{\mathbb{F}} (\tilde{x} , \tilde{t}) =  \tilde{\mathbb{F}}_0 - \tilde{x} \tilde{\phi} (\tilde{t}) $, in which $\tilde{\mathbb{F}}_0 = \tilde{f} - \tilde{\eta}_{\textrm p} \tilde{\phi} ( \tilde{t} )$, where $\tilde{f}$ is the external driving force acting on the monomer(s) inside the nanopore and $\tilde{\eta}_{\textrm p}$ is the pore friction.
Using the fact that the tension force vanishes at the tension front, i.e., $ \tilde{\mathbb{F}} ( \tilde{x} = \tilde{R} , \tilde{t}) = 0 $, $\tilde{R}$ relates to the force at the entrance of the pore on the {\it cis} side as $\tilde{\mathbb{F}}_0 = \tilde{R} (\tilde{t})  \tilde{\phi} ( \tilde{t} )$. Using the definition of the monomer flux in the above relation, the time evolution of the translocation coordinate $\tilde{s}$ can be written as
\begin{equation}
[ \tilde{\eta}_{\textrm p} + \tilde{R} (\tilde{t}) ]  \frac{d \tilde{s}}{ d \tilde{t} } = \tilde{f},
\label{s_Eq}
\end{equation}
where the effective friction inside the brackets is the sum of pore friction ($\tilde{\eta}_{\textrm p}$) and friction due to the movement of the mobile part in the solvent $\tilde{R} (\tilde{t})$. 

To solve Eq.~\!(\ref{s_Eq}), one needs to know the time evolution of $\tilde{R} (\tilde{t})$. By comparing the spatial configuration of the polymer chain at times zero and $\tilde{t}$, the equation of motion for $\tilde{R} (\tilde{t})$ in the TP stage can be written as explained below. As seen in Fig.~\!\ref{fig-schematic}(c), the sum of the number of mobile monomers on the {\it cis} side $\tilde{l}$ and the number of monomers on the {\it trans} side $\tilde{s}$ is denoted by $N$ ($N = \tilde{l} + \tilde{s}$) which is less than the contour length of the polymer ($N  < N_0$) in the TP stage. 
Initially at $t=0$ there is no tension front yet, and thus at time 
 $\tilde{t}$ we can write $\tilde{R}$ as the end-to-end distance of the polymer subchain with contour length $N$ as schematically shown in  Fig.~\!\ref{fig-schematic}(c).

If the channel is wide enough such that $\tilde{D} > \tilde{R}(N_0) = \tilde{R}_{\textrm{e}} $ or we consider early stages of the translocation process during $\tilde{t} < \tilde{\tau}^*$ (panel (b) in Fig.~\!\ref{fig-schematic}) where $\tilde{D} > \tilde{R} (N) $, then $\tilde{R}$ is given by the standard Flory scaling form $\tilde{R} = A  N^{\nu}$. Here, $\nu = 3/4$ is the Flory exponent in two dimensions and $A$ is a constant of order unity. In the high-force limit when the mobile subchain on the {\it cis} side is fully straightened, $\tilde{l}$ is replaced by the distance of the tension front location from the pore, i.e., $\tilde{l} = \tilde{R}$, and using the definition of $N$ the following equation of motion for $\tilde{R}$ must be solved
\begin{equation}
\tilde{R} = A  (\tilde{R} + \tilde{s})^{\nu}.
\label{R_free}
\end{equation}
In the opposite case of narrow channels, in which $\tilde{D} < \tilde{R}(N_0) $, during $\tilde{\tau}^* < \tilde{t} < \tilde{\tau}_\text{TP}$ (panel (c) in Fig.~\!\ref{fig-schematic}) $\tilde{R}$ depends on both $\tilde{D}$ and $N$. For this case, employing the blob theory one can write $\tilde{R}$ as a function of $\tilde{D}$ and $N$.
In Fig.~\!\ref{fig-schematic}(a), the end-to-end distance of the polymer subchain inside each orange blob is written as $\tilde{D} = A g^{\nu}$, where $g$ is the number of monomers inside each blob. The number of blobs inside the channel is $n_{\textrm{b}} = N_0 / g$. Then, at $\tilde{t} = 0$ the end-to-end distance of the chain inside the channel is given by $\tilde{R}_{\textrm{e}} = n_{\textrm{b}} \tilde{D} = A^{1/\nu} N_0 \tilde{D}^{1-1/\nu}$, which is in agreement with the results of the LD simulations (for more details see Appendix~\!\ref{Poly_initial_conf}). Performing the same procedure at time $\tilde{t}$ in panel (c) in Fig.~\!\ref{fig-schematic}, $\tilde{R} (\tilde{t})$ can be obtained as $\tilde{R} = A^{1/\nu} N \tilde{D}^{1-1/\nu}$, where $N = \tilde{l} + \tilde{s}$ and in the high force limit, in which $\tilde{l} = \tilde{R}$, it can be cast to 
\begin{equation}
\tilde{R} = A^{1/\nu} (\tilde{R} + \tilde{s}) \tilde{D}^{1-1/\nu}.
\label{R_channel}
\end{equation}
Thus, in the TP stage the time evolution of $\tilde{R}$ is given by solving either Eq.~\!(\ref{R_free}) or (\ref{R_channel}) in the relevant time regime.

In the PP stage as the tail monomer (chain end) experiences the tension force, the whole {\it cis}-side subchain moves towards the pore. Performing a time derivative of the closure relation in the PP stage, which is $N= \tilde{l} + \tilde{s} = N_0$, in the high force limit ($\tilde{l} = \tilde{R}$), the time evolution of $\tilde{R}$ can be written as
\begin{equation}
\dot{\tilde{R}} = - \tilde{\phi},
\label{R_PP}
\end{equation}
where $\tilde{\phi} = {d \tilde{s}} / {d \tilde{t}}$ is the monomer flux.
To describe the whole translocation process in the regime where the blob theory works, in the TP stage Eqs.~\!(\ref{s_Eq}) and (\ref{R_free}) and (\ref{R_channel}) must be solved, while in the PP stage Eqs.~\!(\ref{s_Eq}) and (\ref{R_PP}) must be considered.

In the case of very narrow channels, in the TP stage the monomers of the mobile part of the chain interact with the particles of the channel walls.
However, in the PP stage the {\it cis}-side mobile part moves like a rod and does not have significant interaction with the particles of the channel walls as long as $D \gg r_{\rm c}$.
Thus, the channel width should not influence the PP stage. However, in
the TP stage the chain-channel interactions induce an additional force $- \tilde{\eta}_{\rm ch} \tilde{\phi} (\tilde{t}) d \tilde{x}'$ on a differential element of the mobile part of the polymer chain on the {\it cis} side, where $\eta_{\rm ch}$ is an effective friction from the walls.  We have numerically evaluated $\eta_{\textrm{ch}} \approx 0.2$ by fitting the waiting time from the IFTP theory on the LD simulation data. Thus, the effective force acting on $d \tilde{x}'$ can be written as $ d \tilde{\mathbb{F}} (\tilde{x}' , \tilde{t}) =  - \tilde{\phi} (\tilde{t}) d \tilde{x}' - \tilde{\eta}_{\rm ch} \tilde{\phi} (\tilde{t}) \Theta ( \tilde{\tau}_{\textrm{TP}} - \tilde{t}) d \tilde{x}' $, where $\Theta$ is the Heaviside step function. Then, the force balance equation for narrow channels is given by
\begin{equation}
\big\{ \tilde{\eta}_{\textrm p} + \tilde{R} (\tilde{t}) [ 1 + \tilde{\eta}_{\textrm{ch}} \Theta ( \tilde{\tau}_{\textrm{TP}} - \tilde{t}) ] \big\}  \frac{d \tilde{s}}{ d \tilde{t} } = \tilde{f}  .
\label{s_Eq_narrow}
\end{equation}
To solve the above equation, $\tilde{R} (\tilde{t})$ must be known. To find an equation for the time evolution of $\tilde{R}$ for narrow channels, in which the blob theory breaks down, the initial configuration of the chain inside the narrow channel and its closure relation must be considered. The end-to-end distance of the polymer chain at $\tilde{t} = 0$ is $\tilde{R}_{\textrm{e}} = B N_0$, where $B \approx 0.77$. Therefore, incorporating the mass conservation $N = \tilde{l} + \tilde{s}$ in the TP stage in the high force limit, in which $\tilde{l} = \tilde{R}$, into $\tilde{R} = B N$, and performing its time derivative, the time evolution of $\tilde{R}$ can be obtained as
\begin{equation}
\dot{\tilde{R}} = \frac{B \tilde{\phi}}{1 - B}.
\label{R_TP_narrow}
\end{equation}
In the PP stage similar to the wider channels the equation of motion for $\tilde{R}$ is simply $\dot{\tilde{R}} = - \tilde{\phi}$.
Therefore, to describe the translocation process for narrow channels in the TP stage Eqs.~\!(\ref{s_Eq_narrow}) and (\ref{R_TP_narrow}) must be solved, while in the PP stage Eq.~\!(\ref{s_Eq_narrow}) and $\dot{\tilde{R}} = - \tilde{\phi}$ must be considered.


\section{Results}\label{results}

In the present work we have systematically studied the influence of the channel width and driving foce on the pore-driven PT process. The values of the channel width are $D=2$, 4, 8, 16, 24, 32, 40, 48, 56, 64, 72, and 128. While the mean and the distribution of the translocation times are investigated for different values of the driving force $f=2$, 4, 8, 16, 20, 40, 60, 80 and 100, other results such as the waiting time (WT) distribution $w$, mean squared displacement (MSD) of monomers, monomer number density $\rho$ and the monomer velocity distributions are investigated in the high force (strong stretching) limit in which the driving force is fixed at $f=100$.


\subsection{Waiting time distribution}

\begin{figure}
\includegraphics[width=0.40\textwidth]{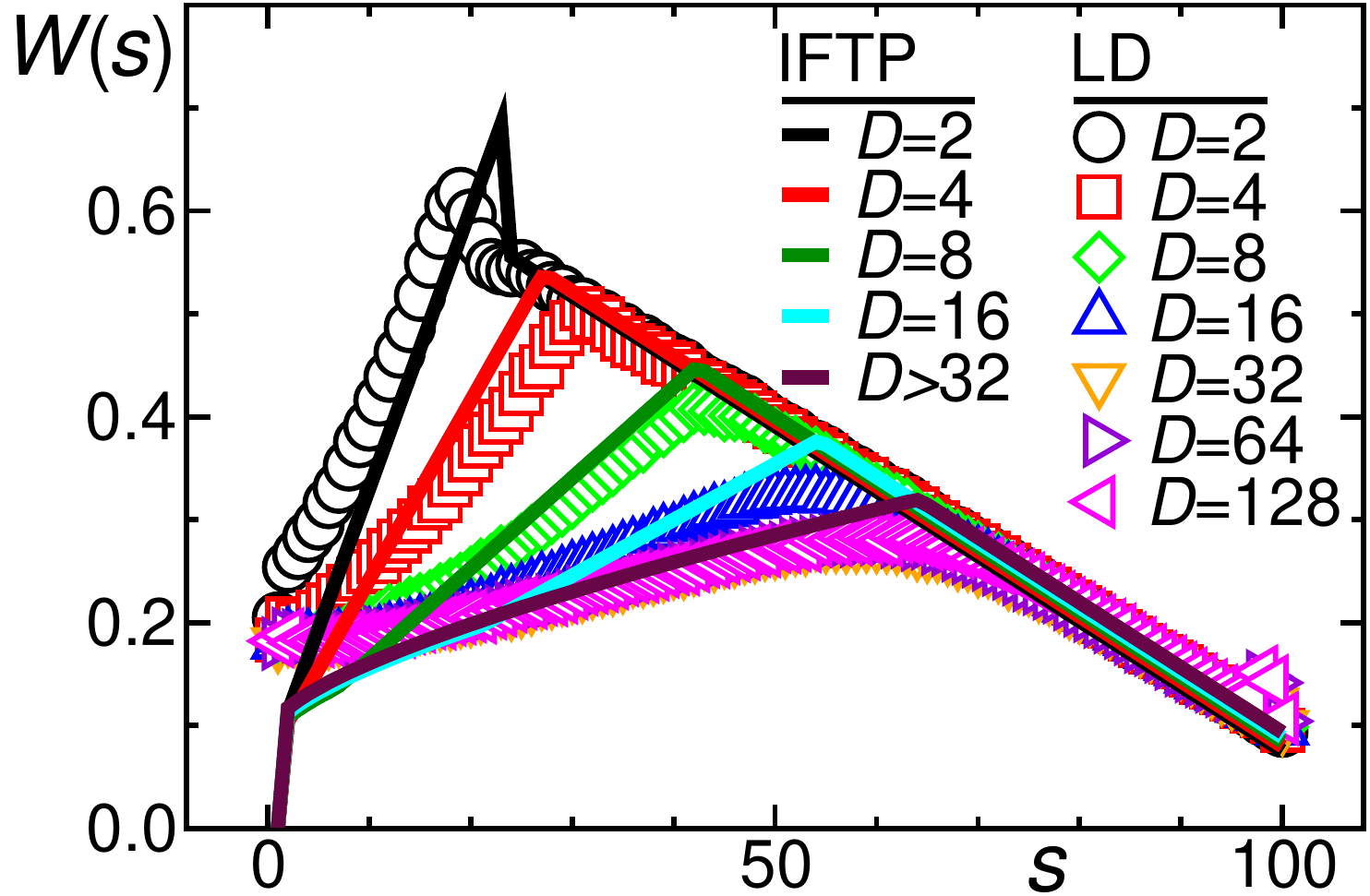}
\caption{Average waiting times $w(s)$ extracted from the LD simulations for $N_0 = 100$ as a function of the translocation coordinate $s$, for fixed value of the driving force $f=100$ and different values of the channel width $D=2$ (black circles), 4 (red squares), 8 (green diamonds), 16 (blue triangles up), 32 (orange triangles down), 64 (violet triangles right) and 128 (magenta triangles left).
The waiting times obtained numerically from the IFTP theory are denoted by the lines for $D=2$ (black line), 4 (red line), 8 (green line), 16 (cyan line), and $D > 32$ (maroon line).
}
\label{fig-Waitings}
\end{figure}

The distribution of waiting times $w(s)$ as a function of the translocation coordinate $s$ is an important quantity that reveals the translocation dynamics of a polymer at the monomer level, and is defined as the average time that each monomer spends at the pore during the translocation process. The waiting time $w(s)$ obtained from the LD simulations is shown in Fig.~\!\ref{fig-Waitings} as a function of the translocation coordinate $s$, for a fixed value of the driving force $f=100$ and different values of the channel width $D=2$ (black circles), 4 (red squares), 8 (green diamonds), 16 (blue triangles up), 32 (orange triangles down), 64 (violet triangles right) and 128 (magenta triangles left). Results from the IFTP theory for $D=2$ (black line), 4 (red line), 8 (green line), 16 (cyan line), and $D > 32$ (maroon line) are in good agreement with the LD simulations. As expected, the pore friction varies with the channel width. The values of pore friction are $\eta_{\rm p}=8$, 8.5, 9, 9.5 and 10, corresponding to the values of the channel width of $D=2$, 4, 8, 16 and $D>32$, respectively. Therefore, the IFTP theory presented in the previous section can successfully explain the local dynamics of the translocation process at the monomer level.

Figure~\!\ref{fig-Waitings} shows that for each $D$, $w(s)$ is nonmonotonic. It first increases and then decreases which is typical for driven PT. In the TP stage ($t < \tau_{\textrm {TP}}$) as time passes more monomers join the mobile subchain on the {\it cis} side and the friction experienced by the chain due to the solvent increases. Consequently, the dynamics slows down and $w(s)$ grows. In the PP stage ($t \geqslant \tau_{\textrm {TP}}$) when the whole {\it cis}-side subchain (influenced by the tension) moves towards the pore, the number of mobile monomers on the {\it cis} side and therefore the solvent friction decrease. Thus $w(s)$ decreases, too.

Moreover, Fig.~\!\ref{fig-Waitings} shows how confinement affects tension propagation. For the smallest value of $D = 2 \ll R_{\textrm e}^{\infty}$, where $R_{\textrm e}^{\infty}\approx 30.3$ 
is the end-to-end distance for a polymer with $N_0=100$ monomers in the semi-infinite free space, the chain fluctuations are strongly suppressed, and tension propagation terminates for the smallest value of $s$ at which the WT maximizes. Increasing $D$ increases the value of $s$ for the maximum of $w(s)$, and finally for $D \ge R_{\textrm e}^{\infty}$ location of the maximum value of $w(s)$ is not affected by further increase in $D$.
%

    
\subsection{Mean translocation time from LD simulations}

\begin{figure}
\includegraphics[width=0.48\textwidth]{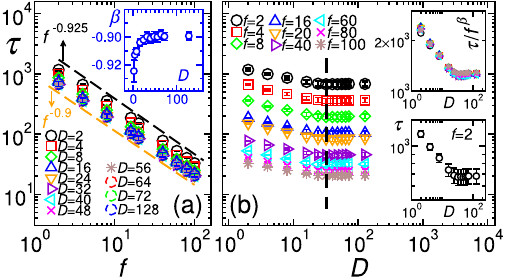}
\caption{(a) The average translocation time $\tau$ from the LD simulations as a function of the driving force $f$ for different values of the channel width $D=2$ (black circles), 4 (red squares), 8 (green diamonds), 16 (blue triangles up), 24 (orange triangles down), 32 (violet triangles right), 40 (cyan triangles left), 48 (magenta crosses), 56 (brown stars), 64 (red dashed circles), 72 (green dashed circles) and 128 (blue dashed circles). The inset shows the force exponent $\beta$ (defined as $\tau\propto f^\beta$) as a function of $D$. The numerically obtained average exponents fall between $-0.925$ and $-0.900$ as indicated in the figure with dashed lines. (b) The average translocation time $\tau$ as a function of the channel width $D$, for different values of the driving force $f=2$ (black circles), 4 (red squares), 8 (green diamonds), 16 (blue triangles up), 20 (orange triangles down), 40 (violet triangles right), 60 (cyan triangles left), 80 (magenta crosses) and 100 (brown stars). The top inset shows the normalized average translocation time $ \tau /f^\beta$ as a function of $D$ for all values of $f$ as in the main panel. For the sake of a better visibility, the bottom inset shows $\tau$ as a function of $D$ for fixed value of $f=2$. The error bars in the top inset in panel (b) are of the order of the symbol sizes or smaller.
}
\label{fig-Tau}
\end{figure}

The mean translocation time $\tau$, which identifies the global dynamics of the translocation process, is the average time that it takes for the whole polymer chain to traverse the pore and move to the {\it trans} side. Figure \ref{fig-Tau}(a) shows the ensemble averaged $\tau$ from LD simulations as a function of the driving force $f$ for different values of the channel width $D=2$ (black circles), 4 (red squares), 8 (green diamonds), 16 (blue triangles up), 24 (orange triangles down), 32 (violet triangles right), 40 (cyan triangles left), 48 (magenta crosses), 56 (brown stars), 64 (red dashed circles), 72 (green dashed circles) and 128 (blue dashed circles). As expected, $\tau$ decreases monotonically with increasing $f$ for all values of $D$. 

The inset in panel (a) shows the force exponent $\beta$ from the scaling form $\tau \propto f^\beta$ as a function of $D$. For the narrowest channel $D=2$ the force exponent is $\beta = -0.925 \pm 0.008$, and it increases with increasing $D$ up to $D \approx R_{\textrm e}^{\infty}$. Then, for $D > R_{\textrm e}^{\infty}$, $\beta$ saturates to $-0.900 \pm 0.003$. The black and orange dashed lines in the main panel in Fig.~\!\ref{fig-Tau}(a) correspond to $f^{-0.925}$ (for $D=2$) and $f^{-0.900}$ (for $D=128$), respectively.

In Fig.~\!\ref{fig-Tau}(b) we plot $\tau$ as a function of the channel width $D$ for different values of the driving force $f=2$ (black circles), 4 (red squares), 8 (green diamonds), 16 (blue triangles up), 20 (orange triangles down), 40 (violet triangles right), 60 (cyan triangles left), 80 (magenta crosses) and 100 (brown stars). The top inset shows the normalized  $\tau /f^\beta$ as a function of $D$ for all values of the driving force in the main panel. The data for $f=2$ are shown in the bottom inset for a better visibility.
The plot shows that for fixed $f$, $\tau$ decreases as $D$ increases up to $D \approx R_{\textrm e}^{\infty}$ (vertical black dashed-line in the main plot). 
This is due to the fact that for small $D$ the chain configuration is more extended along the channel axis and  tension propagates faster, leading to increased drag from the {\it cis}-side subchain.
For $D > R_{\textrm e}^{\infty}$, $\tau$ remains almost constant for fixed $f$. The top right inset in Fig.~\!\ref{fig-Tau}(b) shows data collapse for the normalized $\tau /f^\beta$ as a function of $D$, where we have used the corresponding effective values of $\beta(D)$.

More details on the influence of driving force and channel width on the distribution of the translocation times are given in Appendix~\!\ref{tau_distribution}. 


\subsection{Analytic scaling form of translocation time}

To analytically obtain the scaling form of the translocation time $\tilde{\tau}$ as a function of the chain contour length $N_0$ and channel width $\tilde{D}$ from the IFTP theory, one needs to integrate $N$ from zero to $N_0$ in the TP stage, and $\tilde{R}$ from $\tilde{R} (N_0)$ to zero in the PP stage. 
Tilde indicates dimensionless units, as mentioned in Sec.~\!\ref{theory}. For the cases when the blob theory works, combining the force balance Eq.~\!(\ref{s_Eq}) with the mass conservation laws $N = \tilde{l} + \tilde{s} $ and $N= \tilde{l} + \tilde{s }= N_0$ in the TP and PP stage, respectively, and then summing up the TP and PP times gives
\begin{equation}
\int_{0}^{\tilde{\tau}} \tilde{f} ~ d\tilde{t }=  \int_{N = 0}^{N = N_0 } (\tilde{R} + \tilde{\eta}_{\textrm{p}}) ~ d N.
\label{integral_scaling_time}
\end{equation}
For the case $\tilde{D} < \tilde{R}_{\textrm e}^{\infty}$ the right hand side (RHS) of the above equation is split into different time regimes by comparing $\tilde{D}$ and $\tilde{R}$, and the translocation time can be written as
\begin{equation}
\tilde{\tau}  = \frac{1}{\tilde{f}} \int_{N = 0}^{N = N_*} \tilde{R} ~ dN + \frac{1}{\tilde{f}} \int_{N = N_*}^{N = N_0} \tilde{R} ~ d N + \tilde{\eta}_{\textrm p} N_0 / \tilde{f} ,
\label{integral_scaling_time_2}
\end{equation}
where $N_*$ is obtained as $N_* = (\tilde{D} / A)^{1/\nu}$ by using its definition $\tilde{R} (N_*) = A N_*^{\nu} = \tilde{D}$. In the first term in the RHS of Eq.~\!(\ref{integral_scaling_time_2}) since $\tilde{R} < \tilde{D}$ the integral should be taken over $\tilde{R} (N) = A N^{\nu}$, while in the second term $\tilde{R} > \tilde{D}$ and the channel walls suppress polymer fluctuations, one must use $\tilde{R} (N) = A^{1/\nu} \tilde{D}^{1-1/\nu} N$. Using the definitions of $N_*$ and the proper closure relation for $\tilde{R}$ in Eq.~\!(\ref{integral_scaling_time_2}), the scaling from of the translocation time is obtained as
\begin{equation}
\tilde{\tau}  = \frac{\tilde{D}^{\frac{1+\nu}{\nu}}}{\tilde{f} A^{1/\nu}} \Bigl ( \frac{1}{1+\nu} - \frac{1}{2} \Bigr )
+ \frac{N_0^2}{2 \tilde{f}} A^{1/\nu} \tilde{D}^{1-1/\nu} + \tilde{\eta}_{\textrm p} \frac{N_0}{\tilde{f}}.
\label{scaling_time_3}
\end{equation}
According to this result, for the present case where $\nu=3/4$, $\tau$ decreases with increasing $D$, which is in agreement with our LD data in Fig. \ref{fig-Tau}(b).
A similar equation with the same $N_0$-dependence has already been derived in 
Ref.~\!\onlinecite{Slater_2015}. 

For channels with $\tilde{D} > \tilde{R}_{\textrm e}^{\infty}$, Eq.~\!(\ref{integral_scaling_time}) does not need to be split. Using $\tilde{R} (N) = A N^{\nu}$, the scaling form of the translocation time becomes
\begin{equation}
\tilde{\tau}  = \frac{1}{\tilde{f}} \frac{A}{1+\nu} N_0^{1+\nu} + \tilde{\eta}_{\textrm p} \frac{N_0}{\tilde{f}} ,
\label{scaling_time_3}
\end{equation}
which is identical to that of unconfined pore-driven translocation \cite{jalalJCP2014}.

For a very narrow channel ($D=2$) the blob theory does not work, as mentioned in Sec.~\!\ref{theory}. The result of the IFTP theory (WT) which is confirmed by the results of the LD simulations is shown for $D=2$ in Fig.~\!{\ref{fig-Waitings} (black line and black circles in Fig.~\!{\ref{fig-Waitings}).  
In this case, in the TP stage the mobile monomers on the {\it cis} side experience a dynamical frictional force due to the channel walls, as mentioned in Sec.~\!\ref{theory} to obtain Eq.~\!(\ref{s_Eq_narrow}).
%
%
On the other hand, in the PP stage the chain is almost unaffected by the channel walls as all mobile monomers of the chain in the {\it cis} side move together like a rod towards the pore. 
Therefore, combining the force balance equation $[\tilde{\eta}_{\textrm{p}} + (1 + \tilde{\eta}_{\textrm{ch}}) \tilde{R} ] ({d \tilde{s}}/{d\tilde{t}}) = \tilde{f} $ with  mass conservation $N = \tilde{l} + \tilde{s }$ in the TP stage, and equation $(\tilde{\eta}_{\textrm{p}} + \tilde{R} ) \frac{d \tilde{s}}{d \tilde{t}} = \tilde{f} $ with $N= \tilde{l} + \tilde{s} = N_0$ in the PP stage, and summing up the TP and PP times yields
\begin{eqnarray}
\int_{0}^{\tilde{\tau}} \tilde{f} ~ d \tilde{t} &=&  \int_{N = 0}^{N = N_0 } [ (1+\tilde{\eta}_{\textrm{ch}}) \tilde{R} + \tilde{\eta}_{\textrm p} ] ~ d N \nonumber\\
&-& \int_{\tilde{R} = 0}^{\tilde{R} = \tilde{R}(N_0) } \tilde{\eta}_{\textrm{ch}} \tilde{R} ~ d \tilde{R}. 
\label{integral_scaling_time_1_narrow}
\end{eqnarray}
Here, the effective channel friction $\eta_{\textrm{ch}} = 0.2$, and $\tilde{R}(N) = B N$ with $B \approx 0.77$ is the closure relation for the end-to-end distance of a mobile subchain with $N$ monomers inside a narrow channel.
Using the above closure relation inside Eq.~\!(\ref{integral_scaling_time_1_narrow}), the scaling from of the translocation time is written as 
\begin{equation}
\tilde{\tau} = \frac{1}{\tilde{f}} \big[ \frac{(1 + \tilde{\eta}_{\textrm{ch}})B}{2} - \frac{\tilde{\eta}_{\textrm{ch}} B^2 }{2} \big]  N_0^2 + \frac{1}{\tilde{f}} \tilde{\eta}_{\textrm p} N_0, 
\label{integral_scaling_time_2_narrow}
\end{equation}
where the prefactor ${(1+ \tilde{\eta}_{\textrm{chl}})B}/{2} - {\tilde{\eta}_{\textrm{chl}} B^2}/{2} \approx 0.4$ for the present model.


\subsection{Mean squared displacement}

\begin{figure*}
\includegraphics[width=0.65\textwidth]{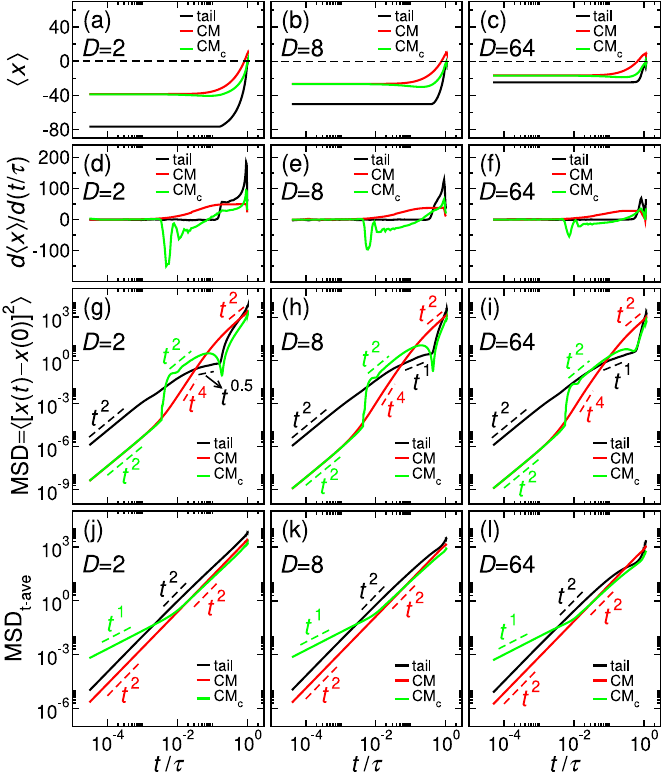} 
\caption{(a) The ensemble average of the horizontal position $\langle x\rangle$ of the tail (black line), center of mass of the whole chain ($\rm CM$) (red line) and center of mass of the subchain in the {\it cis} side ($\text{CM}_{\textrm{c}}$) (green line) as a function of the normalized time $t/\tau$ for fixed values of the driving force $f=100$ and the channel width $D=2$. 
Panels (b)--(c) are the same as (a), but for $D=8$ and 64, respectively. Panels (d)--(f) are the ensemble averages of the slopes $d \langle x \rangle /d(t/\tau )$ corresponding to the curves in panels (a)–(c). (g) Ensemble-averaged MSD as a function of normalized time $t/\tau$ for the tail (black line), CM (red line) and CM$_{{\textrm{c}}}$ (green line) for $f=100$ and $D=2$. Panels (h)--(i) are the same as (g), but for $D=8$ and 64, respectively. (j) Time-averaged MSD, $\rm MSD_{t-ave}$, as a function of $t/\tau$ for the tail (black line), CM (red line) and CM$_{\rm c}$ (green line) for $f=100$ and $D=2$. Panels (k)--(l) are the same as panel (j) but for $D=8$ and 64, respectively. All slopes of the dashed lines in (g)--(i) and (j)--(l) are guides to the eye.
}
\label{fig-Dynamics-x}
\end{figure*}

Next, we study PT by investigating the dynamics of the chain in more detail. To this end, we consider the time-dependence of the mean squared fluctuation of the position $x(t)$ defined by
\begin{equation}
{\rm MSD} = \langle [x(t) - x(0)]^2 \rangle. 
\end{equation}
This quantity is computed for the end monomer (tail) of the chain, the center of mass (CM) of the chain, and the CM of the {\it cis}-side subchain (CM$_{\rm c}$).
In addition, we have computed the 
time-averaged MSD, $\text {MSD}_{\text {t-ave}}$, defined as
\begin{equation}
\text{MSD}_{\text{t-ave}}(j\Delta T)=\frac{1}{n-j}\sum_{i=0}^{n-1-j}\lbrace x[(i+j)\Delta t]-x(i\Delta t)\rbrace^2,
\end{equation}
where $j=$1, 2, 3, ..., $n-1$. In this equation, the $x$ component of the position is saved at times with time lags $\Delta t$ as $x(0)$, $x(\Delta t)$, $x(2\Delta t)$, ..., $x [ (n-1)\Delta t ]$.
 For a non-ergodic system, $\text{MSD}_{\text{t-ave}}$ will be different from the ensemble-averaged MSD \cite{bouchaud,Ralf_Phys_Today,He2008,Jeon2012,Metzler2014}. In an ergodic process the ensemble and the long-time averages are similar.
 
Figures \ref{fig-Dynamics-x}(a)-(c) present ensemble averages of the horizontal position $\langle x \rangle$ of the tail (end monomer) (black line), the CM of the whole chain (red line) and the CM of the {\it cis}-side subchain $\text{CM}_{\rm c}$ (green line) as a function of the normalized time $t/\tau$ for fixed values of the driving force $f=100$ and for different channel widths $D=2,8$ and 64.
The value of $\langle x\rangle$ starts to increase when the tension reaches the tail monomer which for smaller $D$ happens at smaller values of the normalized time.
For all values of $D$, $\langle x \rangle$ of CM (red line) increases as the monomers of the chain are translocated to the {\it trans} side, and eventually becomes positive as the number of the translocated monomers on the {\it trans} side becomes sufficiently large. Moreover, the green lines in panels (a)--(c) correspond to the position of the {\it cis}-side subchain CM, $\text{CM}_{\rm c}$. They show that $\text{CM}_{\rm c}$ first moves away from the pore (as the number of monomers in the mobile domain decreases compared to the number of monomers in the same spatial domain in the equilibrium state at $t=0$), and then it approaches the pore. Finally, at the end of the translocation process at $t / \tau = 1$ the value of $\langle x\rangle$ for $\text{CM}_{\rm c}$ becomes zero.

Panels (d)--(f) present the slopes $d \langle x \rangle / d (t/\tau)$ corresponding to the curves in panels (a)--(c). As can be seen, the slope for the position of the tail monomer (in black color) starts to increase when the tension reaches the tail monomer which for smaller $D$ happens at smaller values of the normalized time. The slope for the position of $\text{CM}_{\rm c}$ (in green color) at short times is zero and then becomes negative as the location of  $\text{CM}_{\rm c}$ moves towards the pore. As time passes and the $\text{CM}_{\rm c}$ approaches the pore, the slopes become positive. The slope for the position of CM (in red color) is zero at early times and positive at the intermediate and long time regions.

\begin{figure*}
\includegraphics[width=0.75\textwidth]{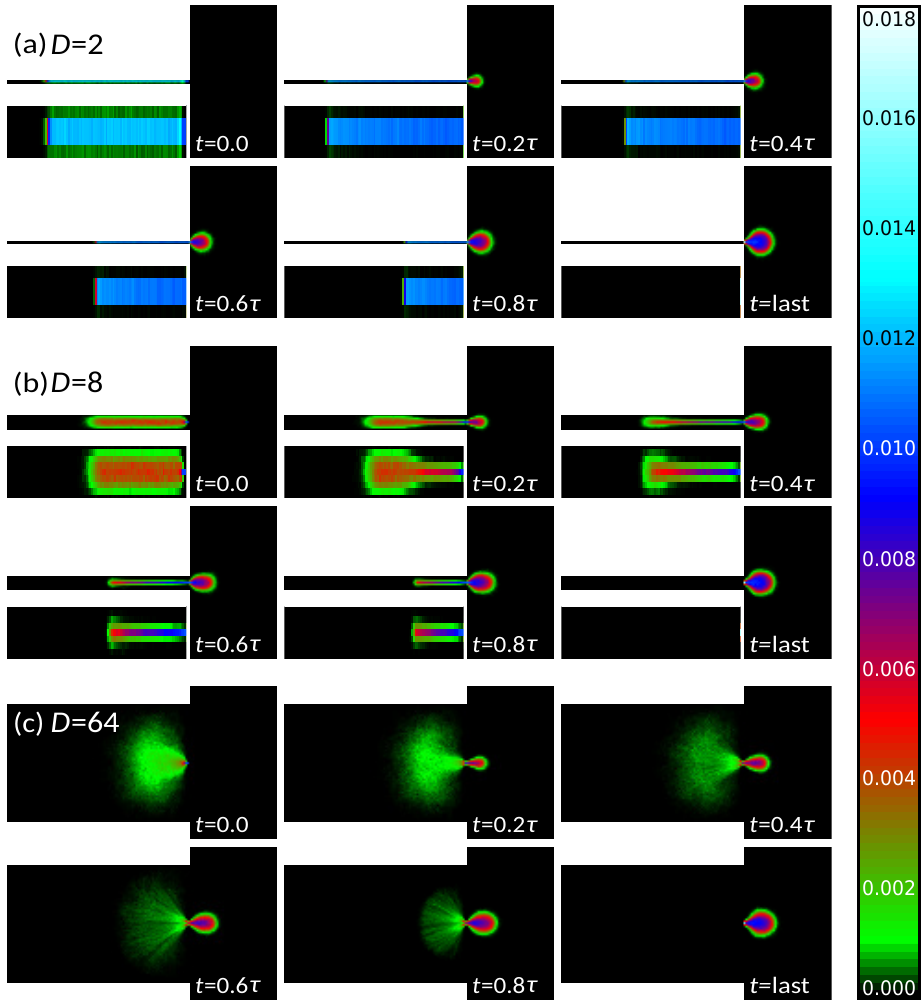}
\caption{(a) Monomer number density $\rho(x,y)$ for fixed values of the driving force $f=100$ and the channel width $D=2$ at different times $t/\tau=0$, 0.2, 0.4, 0.6, 0.8 and at the last snapshot. Panels (b) and (c) are the same as panel (a), but for other values of the channel width $D=8$ and 64, respectively. For the sake of better visibility, $\rho(x,y)$ of the {\it cis} side (channel) is magnified and shown below each actual channel in panels (a)--(b).
}
\label{fig-Density}
\end{figure*}


In Fig.~\ref{fig-Dynamics-x}(g) the ensemble-averaged MSD is shown as a function of $t/\tau$ for the tail (black line), CM (red line) and CM$_{\rm c}$ (green line) for $f=100$ and $D=2$. Panels (h)--(i) are the same as panel (g) but for $D=8$ and 64, respectively.
In all time regimes the behavior of MSDs for the tail (and also for the $\rm CM$) is similar for different values of $D$. The MSDs for the tail monomer (black lines in panels (g)--(i)) scale as $t^2$ at early times and then show more complicated behavior when the PP stage starts.
In contrast, MSDs for CM (red lines in panels (g)--(i)) show more smooth behavior, scaling intially also as $t^2$. In addition, the MSDs for CM$_{\rm c}$ (green lines in panels (g)--(i)) scale again as $t^2$ at early times but then follow the black curves at late times. The sharp valleys in the green curves represent the transition from TP to PP.


In Fig. \ref{fig-Dynamics-x}(j) $\text{MSD}_{\text{t-ave}}$ (averaged over 75 independent trajectories) is shown as a function of the normalized time $t/\tau$ for the tail (black line), $\rm CM$ (red line) and $\text{CM}_{\rm c}$ (green line) for fixed values of the driving force $f=100$ and the channel width $D=2$. Panels (k)--(l) are the same as panel (j) but for $D=8$ and 64, respectively. 
As panels (j)--(l) clearly show, $\text{MSD}_{\text{t-ave}}$ for the tail and CM scales as $t^2$ for almost the entire translocation process for all values of $D$. On the other hand, the slope of $\text{MSD}_{\text{t-ave}}$ of the CM$_{\textrm{c}}$ (green line) crosses over from linear dependence to $t^2$. At very late times for $D=8$ and 64 there is a slight bend in the tail curve indicating transition 
from TP to PP.

%
%



\subsection{Monomer Density}

Finally, 
we study the time evolution of the monomer number density $\rho(x,y)$ (the number of monomers per unit area), where the average spatial configuration of the polymer is investigated as a function of time during the translocation process. To calculate $\rho$, the 2D simulation box is divided into square unit cells of size $1\times 1$ each. The number of monomers in each unit cell is counted at a given moment for each trajectory, and divided by the contour length of the polymer. At the end, the averaging is done over 1000 independent trajectories.
Figure \ref{fig-Density}(a) shows the monomer number density at different normalized times $t/\tau=0$, 0.2, 0.4, 0.6, 0.8 and 1.0 for fixed values of the driving force $f=100$ and channel width $D=2$. Panels (b) and (c) are the same as panel (a), but for different values of $D=8$ and 64, respectively. For the sake of better visibility, the monomer number density of the {\it cis} side (inside the channel) is magnified and separately shown below the actual channel in panels (a)--(b). 

As can be seen in panel (a) of Fig.~\!\ref{fig-Density}, in this case the polymer translocates in a rodlike manner in the channel forming a coil at the {\it trans} side. 
Figure \ref{fig-Density}(b) shows that increasing the channel width to $D=8$ causes the chain to extend in the $y$ direction until the tension front reaches the end. 
Finally, 
panel (c) shows that for $D=64$ there is no contact between the chain and the particles on the channel walls since $D > R_{\textrm e}^{\infty}$.


\section{Summary and Conclusions}\label{conclusion}

In this work, we have revisited the problem of pore-driven PT from a confining channel into semi-infinite free space. To this end, we have employed the IFTP theory to analytically calculate the scaling form for the average translocation time as a function of chain length, driving force, confinement, and channel friction. The IFTP theory can quantitatively explain the local waiting times as well as global dynamics as given by the translocation time for the PT process for different values of channel width. The theory includes all cases from a highly confining channel where the blob theory breaks down, to intermediate channel widths where the blob theory applies, and finally to wide channels where spatial confinement becomes irrelevant.

First, for highly confining channels with $D \ll {R_{\textrm e}^{\infty}}$ comparison between the waiting time from LD simulations and from the IFTP theory reveals that in the TP stage interactions between the mobile monomers with the channel walls induce dynamic friction that is proportional to the size of the mobile domain at the {\it cis} side.  In the PP stage the mobile monomers move in a rodlike fashion inside the channel and do significantly interact with the channel walls. Consequently, IFTP theory shows that the translocation time is similar to that of a rod and scales as $\tilde{\tau} \propto N_0^2 $~\!\cite{jalalSciRep2017}.

Second, for a channel of an intermediate width $1\ll D < {R_{\textrm e}^{\infty}}$}, we have used the blob theory in conjunction with the IFTP theory. The predictions on local and global PT dynamics are again in full agreement with the LD simulations. The scaling form of the translocation time depends on both the chain contour length and the channel diameter and agree with those in  Ref. \onlinecite{Slater_2015} in the appropriate limits.

Third, for wide channels with $D > R_{\textrm e}^{\infty}$, both} the IFTP theory and LD simulations  recover the well-known results for a system with a semi-infinite space on the {\it cis} side, where the IFTP theory shows that the translocation time scales as $\tilde{\tau} \propto N_0^{1+\nu} $~\!\cite{jalalJCP2014}.

Finally, we have further characterized the details of the local and global chain dynamics by investigating the distribution of translocation times (cf. Appendix B) and the dependence of the MSDs of tail, CM and CM of the {\it cis}-side subchain, time evolution of the monomer number density and monomer velocity (cf.  Appendix~\!\ref{mon_velocity}) on the channel width at constant driving force in the high force limit. Combined together, our results here give a complete picture on the PT dynamics from confined channels to free space and can be used to interpret and analyze related experiments of PT dynamics and biopolymer sequencing with nanopores.


\appendix

\section{Polymer scaling from blob theory in a channel}
\label{Poly_initial_conf}

To study the scaling of the polymer chain in a channel, we consider the radius of gyration (RG) $R_{\text g}$ and end-to-end distance $R_{\text e}$. The ensemble averages $\langle ... \rangle$ of $R_{\text g}$ and $R_{\text e}$ for a polymer chain with contour length of $N$ are given by~\!\cite{Rubinstein} $R_{\text g}^2 = \frac{1}{N}\sum_{i=1}^{N}\langle(\vec{R}_i-\vec{R}_{\text{CM}})^2\rangle$ and $R_{\text e}^2 = \langle( \vec{R}_N-\vec{R}_1 )^2\rangle$, respectively, 
where $\vec{R}_{i}$ and $\vec{R}_{\text{CM}}$ are the position vectors of the $i$th monomer and the center of mass of the chain, respectively. 

\begin{figure}
\includegraphics[width=0.48\textwidth]{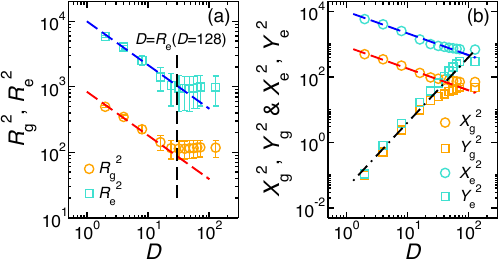}
\caption{
(a) The average squared gyration radius $R_{\text g}^2$ (orange circles) and the average squared end-to-end distance $R_{\text e}^2$ (cyan squares) of the initial configuration of the chain at $t=0$ inside the channel as a function of the channel width $D$. The blue and the red dashed lines represent $(100\times D^{1-\nu})^2$ and $(29\times D^{1-\nu})^2$ with $\nu=3/4$, respectively. The black vertical dashed line is $D=30.32$. 
(b) The average squared radius of gyration in the $x$ and $y$ directions $X_{\text g}^2$ (orange circles) and $Y_{\text g}^2$ (orange squares), and the average squared end-to-end distance in the $x$ and $y$ directions $X_{\text e}^2$ (cyan circles) and $Y_{\text e}^2$ (cyan squares) of the initial configuration at $t=0$, respectively, as a function of the channel width $D$. The blue and the red dashed lines are the same as of panel (a). The black dashed-dotted line is $(0.2\times D)^2$.
}
\label{fig-Rg-Re}
\end{figure}

\begin{figure*}
\includegraphics[width=0.75\textwidth]{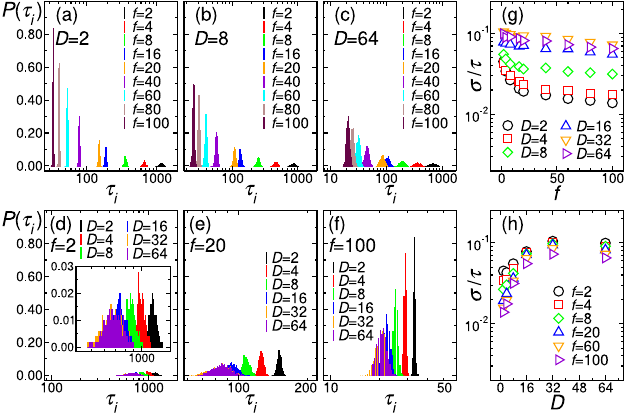}
\caption{
(a) Probability density function of the translocation times $P(\tau_i)$ as a function of the translocation time $\tau_i$, for fixed value of the channel width $D=2$ and different values of the driving force $f=2$ (black bars), 4 (red bars), 8 (green bars), 16 (blue bars), 20 (orange bars), 40 (violet bars), 60 (cyan bars), 80 (brown bars) and 100 (maroon bars). Panels (b)--(c) are the same as panel (a) but for other values of the channel width $D =8$ and 64, respectively. 
(d) $P(\tau_i)$ as a function of $\tau_i$, for fixed value of the driving force $f=2$ and different values of the channel width $D=2$ (black bars), 4 (red bars), 8 (green bars), 16 (blue bars), 32 (orange bars) and 64 (violet bars). Panels (e)--(f) are the same as panel (d) but for different values of the driving force $f = 20$ and 100, respectively. The inset in panel (d) corresponds to the magnified main panel for better visibility.
(g) Normalized value of the standard deviation $\sigma/\tau$ as a function of $f$, for different values of the channel width $D=2$ (black circles), 4 (red squares), 8 (green diamonds), 16 (blue triangles up), 32 (orange triangles down) and 64 (violet triangles right).
(h) $\sigma/\tau$ as a function of $D$, for different values of the driving force $f=2$ (black circles), 4 (red squares), 8 (green diamonds), 20 (blue triangles up), 60 (orange triangles down) and 100 (violet triangles right).
}
\label{fig-Histo}
\end{figure*}

As mentioned in Sec.~\!\ref{theory}, using the blob theory the end-to-end distance of a chain confined in a channel with diameter of $D$ is given by $R_{\text e}=N A^{1/\nu} D^{1-1/\nu}$.
Figure~\!\ref{fig-Rg-Re}(a) compares the results of LD simulations with those of the blob theory for $R_{\text g}$ and $R_{\text e}$. In panel (a) of Fig.~\!\ref{fig-Rg-Re}, $R_{\text g}^2$ (orange circles) and $R_{\text e}^2$ (cyan squares) of the initial configuration of the polymer inside the channel at $t=0$ are plotted as a function of the channel width $D$.
As can be seen, as $D$ increases, the values of both $R_{\text g}^2$ and $R_{\text e}^2$ decrease and in the regime in which $D > R_{\text e} (D=128) \approx  30.32$, they remain almost constant.
The blue and red fitted dashed lines represent $(100\times D^{1-\nu})^2$ and $(29\times D^{1-\nu})^2$ with $\nu =3/4$, respectively. 
Therefore, we conclude that the results of LD simulations are in very good agreement with the blob theory.

Panel (b) in Fig. \ref{fig-Rg-Re} shows the average of the square of the RG and the end-to-end distance in the $x$ and $y$ directions as $X_{\text g}^2$ (orange circles) and $Y_{\text g}^2$ (orange squares), and as $X_{\text e}^2$ (cyan circles) and $Y_{\text e}^2$ (cyan squares), respectively, for the initial ($t=0$) configuration of the chain inside the channel as a function of $D$. The blue and red dashed lines are the same as those in panel (a). The black dashed-dotted line is $(0.2\times D)^2$ and is obtained by fitting to the data in the regime of $D < R_{\textrm{e}} (D=128)$.


\section{Distribution of translocation times}
\label{tau_distribution}

In this Appendix, we present LD results for the probability density function of translocation times $P(\tau_i)$. In Fig.~\!\ref{fig-Histo}(a) $P(\tau_i)$ is shown as a function of individual translocation times $\tau_i$ for fixed value of the channel width $D=2$ and different values of the driving force $f=2$ (black bars), 4 (red bars), 8 (green bars), 16 (blue bars), 20 (orange bars), 40 (violet bars), 60 (cyan bars), 80 (brown bars) and 100 (maroon bars). Panels (b)--(c) are the same as panel (a) but for other values of the channel width $D =8$ and 64, respectively. For fixed $D$, increasing $f$ obviously decreases the average $\tau$ and $P(\tau_i)$ becomes narrower. In fact, as $f$ decreases,, the spatial fluctuations of the configurations of the {\it cis}-side subchain increase significantly, so the fluctuations of the translocation times also increase.

\begin{figure*}
\includegraphics[width=0.9\textwidth]{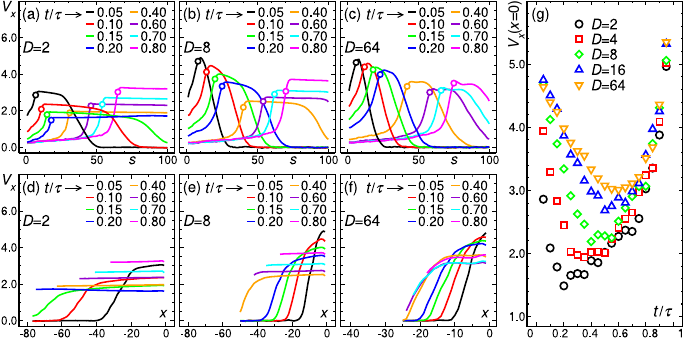}
\caption{
(a) The $x$ component of the monomer average velocity $v_x$ as a function of the translocation coordinate $s$ for fixed values of $D=2$ and $f=100$ at different times $t/\tau=0.05$ (black line), 0.10 (red line), 0.15 (green line), 0.20 (blue line), 0.40 (orange line), 0.60 (violet line), 0.70 (cyan line) and 0.80 (magenta line). Panels (b)--(c) are the same as panel (a), but for other values of the channel width $D=8$ and 64, respectively. The open circle on each curve shows the mean value of the index of the monomer inside the pore at the corresponding time. 
(d) The $x$ component of the monomer average velocity $v_x$ as a function of $x$ in the {\it cis} side, for fixed values of $D=2$ and $f=100$ at different times $t/\tau=0.05$ (black line), 0.10 (red line), 0.15 (green line), 0.20 (blue line), 0.40 (orange line), 0.60 (violet line), 0.70 (cyan line) and 0.80 (magenta line). Here $x=0$ is the location of the pore. Panels (e)--(f) are the same as panel (d), but for other values of the channel width $D=8$ and 64, respectively. Here, $|x|$ is the distance from the pore on the {\it cis} side.
(g) The average velocity of monomer(s) inside the pore $v_x (x=0)$ as a function of $t/\tau$ for fixed value of driving force $f=100$ and different values of the channel width $D=2$ (black circles), 4 (red squares), 8 (green diamonds), 16 (blue triangles up) and 64 (orange triangles down).
}
\label{fig-Vx}
\end{figure*}

In Fig.~\!\ref{fig-Histo}(d) $P(\tau_i)$ is shown as a function of translocation time $\tau_i$, for fixed value of the driving force $f=2$ and different values of the channel width $D=2$ (black bars), 4 (red bars), 8 (green bars), 16 (blue bars), 32 (orange bars) and 64 (violet bars). Panels (e)--(f) are the same as panel (d) but for other values of the driving force $f = 20$ and 100, respectively. For the sake of better visibility, the inset in panel (d) shows the magnified main panel. 
As can be seen for smaller values of $D$, increasing the value of $f$ leads to an increase in the separation of $P(\tau_i)$'s, while this is not the case for larger values of $D$, in which for different values of $f$ the probability density functions show an overlap.   

In Fig.~\!\ref{fig-Histo}(g) the normalized standard deviation of translocation times $\sigma/\tau$ is shown as a function of $f$, for different values of the channel width $D=2$ (black circles), 4 (red squares), 8 (green diamonds), 16 (blue triangles up), 32 (orange triangles down) and 64 (violet triangles right).  
It is clear that $\sigma/\tau$ decreases monotonically as $f$ increases for all values of $D$, and the reduction of $\sigma/\tau$ with respect to $f$ is more pronounced for smaller values of $D$.

Moreover, $\sigma/\tau$ is displayed in panel (h) as a function of $D$, for different values of the driving force $f=2$ (black circles), 4 (red squares), 8 (green diamonds), 20 (blue triangles up), 60 (orange triangles down) and 100 (violet triangles right). 
As can be seen, for each value of $f$, $\sigma/\tau$ increases as $D$ increases. At constant $D$ the deviation of $\sigma/\tau$ with respect to $f$ is more pronounced for {larger} values of $D$ than smaller ones.


\section{Monomer velocities for the translocation coordinate}
\label{mon_velocity}

Here the dynamics of the {\it cis}-side subchain and the tension front are investigated by studying the velocity distribution of each individual monomer at different moments during the translocation process. In Fig.~\!\ref{fig-Vx}(a) the $x$ component of the monomer velocity $v_x$ is plotted as a function of the translocation coordinate $s$ for fixed values of $D=2$ and $f=100$ at different times $t/\tau=0.05$ (black line), 0.10 (red line), 0.15 (green line), 0.20 (blue line), 0.40 (orange line), 0.60 (violet line), 0.70 (cyan line) and 0.80 (magenta line). Panels (b)--(c) are the same as panel (a), but for other values of the channel width $D=8$ and 64, respectively. 
In each curve, the left side of the open circle corresponds to the velocity of the translocated monomers to the {\it trans} side, and its right side shows the velocity of the monomers on the {\it cis} side, at the corresponding time. 

After equilibration at $t=0$ all monomers have zero mean velocity as monomers have not experienced any tension yet.
As can be seen in Fig.~\!\ref{fig-Vx} for each value of $D$, the number of monomers with non-zero velocity increases as time passes, i.e., more monomers join the mobile part of the chain on the {\it cis} side. Moreover, for smaller values of $D$ the last monomer feels the tension (joins to the mobile part) at smaller values of $t/\tau$. Therefore, for smaller values of $D$ the PP stage has a more significant contribution to the mean translocation time than the TP stage. 
Conversely, in the case where the channel width has larger values, i.e., $D > R_{\textrm e}^{\infty}$, it takes  longer for the tension to reach the last monomer and therefore the TP stage has larger contribution to the mean translocation time than the PP stage.
In addition, panels (a)--(c) show that in the TP stage ($t<\tau_{\text{TP}}$) the velocity of the mobile part of the chain on the {\it cis} side decreases with time, originating from the fact that as time passes more monomers join the mobile part and the effective friction increases slowing down the dynamics. Conversely, in the PP stage ($t > \tau_{\text{TP}}$) the velocity of the mobile monomers on the {\it cis} side increases with time due to the reduction in friction.

Furthermore, in Fig.~\!\ref{fig-Vx}(d) the mean value of the $x$ component of the monomer velocity $v_x$ is shown as a function of the $x$ coordinate on the {\it cis} side, for fixed values of $D=2$ and $f=100$ at different times $t/\tau=0.05$ (black line), 0.10 (red line), 0.15 (green line), 0.20 (blue line), 0.40 (orange line), 0.60 (violet line), 0.70 (cyan line) and 0.80 (magenta line). Panels (e)--(f) are the same as panel (d), but for other values of the channel width $D=8$ and 64, respectively. 
To find $v_x$ as a function of $x$, the {\it cis}-side channel is divided into parallel bins and $x$ denotes the location of the bins. $x=0$ indicates location of the pore. The absolute value of $x$ identifies the distance of the corresponding bin on the {\it cis} side from the pore.

In panel (g) the average velocity of monomer(s) inside the pore $v_x (x =0)$ is plotted as a function of the normalized time $t/\tau$ for fixed value of driving force $f=100$ and various values of $D=2$ (black circles), 4 (red squares), 8 (green diamonds), 16 (blue triangles up) and 64 (orange triangles down). For each $D$ as time passes $v_x (x =0)$ first decreases (in the TP stage where $t < \tau_{\textrm{TP}}$), attains a minimum (transition from TP to the PP stage) and then increases (in the PP stage where $t > \tau_{\textrm{TP}}$).
As can be seen, in the TP stage and at a constant time increasing $D$ increases the velocity of monomers at the pore increases. 
In contrast, in the PP stage as time passes the velocity increases due to the reduction in the number of mobile monomers on the {\it cis} side. Moreover, in the PP stage as time passes the velocities eventually collapse on a master curve. The results in panel (g) are in agreement with the waiting times presented in Fig.~\!\ref{fig-Waitings}.


\begin{acknowledgements}
SE and JS acknowledge Iran National Science Fundation: "This work is based upon research funded by Iran National Science Fundation (INSF) under project No.~\!4026895". RM acknowledges the German Science Foundation (DFG, grants ME 1535/16-1 and ME 1535/13-1) for support. T.A-N. has been supported in part by the Academy of Finland grant no. 353298 under the European Union – NextGenerationEU instrument.\end{acknowledgements}




\end{document}